\documentclass[aps,prl,10pt,preprint,nofootinbib]{revtex4-1}
\usepackage{graphicx}
\usepackage{amsmath}
\usepackage{amssymb}
\usepackage{color}
\begin{document}

\title{High harmonic generation with circularly polarized fields in solid: a quantum trajectory perspective}

\author{Liang Li$^{1}$, Pengfei Lan$^{1}$}
\email{pengfeilan@hust.edu.cn}
\author{Xiaosong Zhu}
\email{zhuxiaosong@hust.edu.cn}
\author{Tengfei Huang}
\author{Peixiang Lu$^{1,2}$}
\email{lupeixiang@hust.edu.cn}

\affiliation{%
 $^1$Wuhan National Laboratory for Optoelectronics and School of Physics, Huazhong University of Science and Technology, Wuhan 430074, China\\
$^2$Laboratory of Optical Information Technology, Wuhan Institute of Technology, Wuhan 430205, China
}%

\date{\today}

\begin{abstract}
We investigate the high harmonic generation (HHG) in solids driven by laser fields with different ellipticities. The HHG spectra show a two-plateau structure within the energy gap between the valence band and the first conduction band. These two plateaus depend distinctly on the laser ellipticity. The first plateau is decreased while the second plateau is enhanced with increasing the laser ellipticity. To understand these phenomena, we develop an intuitive Reciprocal-Space-Trajectory (RST) method, with which HHG in solids is explained by a trajectory-ensemble from different initial states and different ionization times in the reciprocal space. In the framework of RST, we can not only quantitatively reproduce the HHG spectra, but also well understand the underlying physics of these phenomena, providing a deep insight into the mechanism of HHG in solids.
\end{abstract}                         
\maketitle

High harmonic generation (HHG) from gas-phase atoms and molecules has opened up a new frontier in ultrafast science, where the generation of attosecond pulses \cite{Hentschel,Paul} and the measurement with attosecond temporal and Angstrom spatial resolutions become accessible \cite{Uiberacker2007,Baker,Blaga,Pertot,LixinHe2018}. In the very recent years, HHG has also been observed in solids \cite{Ghimire2011,TTLuu2015,Ghimire2016}, which makes it possible to extend the successful attosecond  metrology to solid phase systems.

The HHG in two-dimensional (2D) non-trivially polarized laser fields in gases has attracted a great deal of attentions in the past years, because it will ignite a deeper insight into the mechanism of HHG \cite{Burnett1995,Budil1993} and also promise unprecedented applications, such as generation of circularly polarized high harmonics \cite{circular1,circular2} and uncovering the ultrafast dynamics in atoms \cite{chiral}, etc. Accordingly, the HHG in solids with the circularly or elliptically polarized laser fields has attracted close attentions very recently \cite{YouYS2016,Ndabashimiye,Dejean2017,Yoshikawa2017,Corkum2015}. However, because the solid systems have much more complex structures and dynamical processes than the gaseous medium, the HHG in solids exhibit many distinct and unexplored features, e.g., see Refs. \cite{YouYS2016,Yoshikawa2017,Ndabashimiye}. Although several numerical models, such as numerically solving time-dependent Sch\"{o}dinger equation (TDSE) \cite{MW2015,Bian2017,Takuya2017,Lewenstein2017}, semiconductor Bloch equation (SBE) \cite{Koch2003,Golde2008,Vampa2014,TTLuu2016}, and time-dependent density functional theory (TDDFT) \cite{Runge1984,Leeuwen1998} can give good descriptions of HHG, the underlying mechanisms are buried in the wave functions. A generalized re-collision model \cite{Vampa2014,Vampa2015} is also proposed following the counterpart of HHG in gases \cite{Corkum1993}. It can provide some intuitive explanation \cite{MW2015,Bian2017} but fails to give a satisfactory description in quantity. Moreover, due to the complexity of the HHG in solids, many phenomena, especially under the non-trivially polarized laser fields, still cannot be well understood.



In this Letter, we investigate the HHG in solids driven by laser fields with different ellipticities. We find that the HHG spectra show two plateaus within the energy gap between the valence band (VB) and the first conduction band (CB). More importantly, the ellipticity dependencies of these two plateaus are opposite. To explain these phenomena, we develop an intuitive Reciprocal-Space-Trajectory (RST) method based on accelerated-Bloch states, in which the HHG can be explained in terms of the quantum path integral of the trajectory-ensemble. Our RST method fully includes quantum interference effect and allows us to reproduce the HHG spectrum. With this method, the origin of the novel two-plateau structure and the abnormal ellipticity dependencies can be intuitively understood.

\begin{figure}[!t]
  \includegraphics[width=10cm]{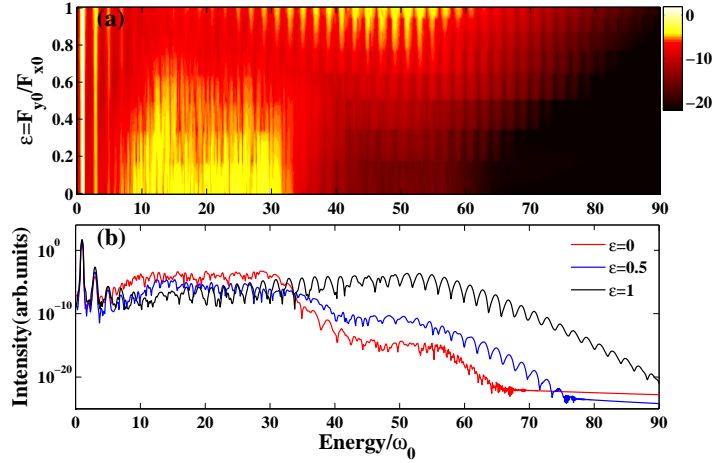}
  \caption{(a) High harmonic spectra for different ellipticities $\varepsilon = F_{y0}/F_{x0}$. (b) High harmonic spectra for $\varepsilon=$ 0, 0.5 and 1.}\label{fig1}
\end{figure}

Figure \ref{fig1} shows the high harmonic spectra of ZnO driven by laser pulses with different ellipticities. The harmonic spectra are obtained by numerically solving the SBE \cite{Vampa2014}. The dephasing effect does not change the general structure of the harmonic spectra as shown in Ref. \cite{Vampa2014}, and is neglected in this work. For computational convenience, we perform the two-dimensional(2D) calculations (i.e., $k_z=0$ in the reciprocal-space) in the following discussions. The 2D calculation can well reproduce the features of HHG in the 3D simulation (see the Supplemental Material \cite{SM}). We use the same band structure as in Ref. \cite{Vampa2014}. The frequency of the driving field is $\omega = 0.014$ atomic units (a.u.). The laser pulse is polarized in the x-y plane, and the ellipticity is varied by keeping the laser amplitude in the major axis constant ($F_{x0}=0.004$ a.u.) and varying the amplitude in the minor axis ($F_{y0}$) from 0 to $0.004$ a.u.. The electric field oscillates under a sine squared envelope with the duration of 10 optical cycles. As shown in Fig. \ref{fig1}, the harmonic spectra firstly decrease with the harmonic order, and then exhibit a two-plateau structure. The first plateau is from the 10{-th} to 29{-th} harmonics, and the second plateau is from the 40{-th} to 55-th harmonics. This two-plateau structure is different from that resulting from the multiple conduction bands \cite{MW2015}, because the spectral range is within the energy gap between the VB and the first CB.
In the linearly polarized field, the first plateau is much stronger than the second plateau and has a clear cutoff at the 29{-th} harmonic. In the elliptically polarized field, the first plateau becomes weaker while the second plateau becomes stronger. In the circularly polarized field, the second plateau becomes even stronger than the first plateau, and the two plateaus merge. In addition, the harmonic yield in a circularly polarized field is comparable to or even higher than that in the linearly polarized field, and the cutoff energy is also larger in the circularly polarized field. These features are very different from those in gaseous mediums, where almost no harmonics are generated with a circularly polarized field.

To analyze the HHG process in solids and understand these phenomena, we develop an intuitive RST method. Within this method, the HHG in solids is described by the coherent buildup of a trajectory-ensemble from different initial states and different ionization times in the reciprocal space. In the reciprocal space, one can restrict the periodical solid system in the first Brillouin zone (BZ) according to the Bloch theorem. In the present of a homogeneous external field, the instantaneous eigenstate $|\varphi(t)\rangle$ of Hamiltonian $\hat{H}(t)$ with eigenenergy $\tilde{E}(t)$ in velocity-gauge satisfies:
\begin{align}\label{Hamiltonian}
  \hat{H}(t)|\varphi(t)\rangle = (\frac{[\hat{\textbf{p}}+\textbf{A}(t)]^{2}}{2}+U(\hat{\textbf{r}}))|\varphi(t)\rangle = \tilde{E}(t)|\varphi(t)\rangle,
\end{align}
where the periodicity of the Hamiltonian is maintained. $\hat{\mathbf{p}}$ is the momentum operator, $U(\hat{\textbf{r}})$ is the Coulomb potential, and $\mathbf{A}(t)$ is the vector potential of the laser field. The requirement that the solutions of Eq. \ref{Hamiltonian} satisfy Bloch theorem and the Born-von K\'{a}rm\'{a}n boundary conditions yields \cite{Krieger1986}
\begin{align}\label{eigenstates}
  \langle\textbf{r}|\varphi_{m,\textbf{k}_{0}}(t)\rangle &= e^{-i\textbf{A}(t)\cdot\textbf{r}}\phi_{m,\textbf{k}(t)}(\textbf{r}), \nonumber \\
  \tilde{E}_{m,\textbf{k}_{0}}(t) &= E_{m}(\textbf{k}(t)),
\end{align}
where $\textbf{k}(t) = \textbf{k}_{0}+\textbf{A}(t)$ is the time-dependent crystal momentum satisfying the acceleration theorem $d\textbf{k}(t)/dt = -\textbf{F}(t) = d\textbf{A}/dt$.  $\textbf{F}(t)$ is the electric field. $\textbf{k}_0$ is the crystal momentum of the electron at the initial time $t_0$. $\phi_{m,\textbf{k}}(\textbf{r})$ is the Bloch state and $E_{m,\textbf{k}}$ is the energy corresponding to the band index $m$ and crystal momentum $\textbf{k}$. The states $|\varphi_{m,\textbf{k}_{0}}(t)\rangle$ are the so-called accelerated-Bloch states \cite{Yakoelev2016}.

\begin{figure}[!t]
  \includegraphics[width=10cm]{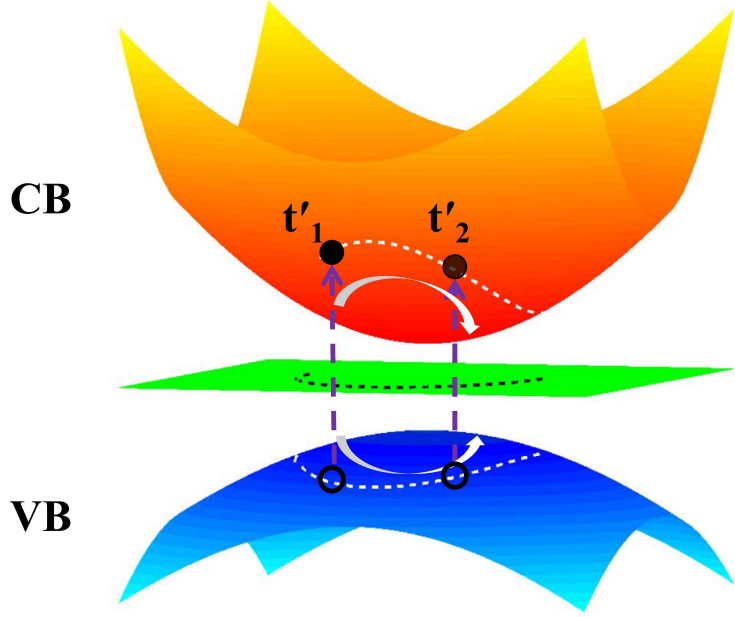}
  \caption{The sketch of the trajectories in reciprocal space. The white dashed lines mark the oscillations in VB and CB.}\label{fig2}
\end{figure}

Based on the accelerated-Bloch states, the evolution of the wave packet of the electron can be described as \cite{Yakoelev2016}
\begin{align}\label{wave function}
  |\Psi_{\textbf{k}_{0}}(t)\rangle = \sum_{m}\alpha_{m,\textbf{k}_{0}}(t)e^{-i\int_{t_{0}}^{t}d\tau E_{m}(\textbf{k}(\tau))}|\varphi_{m,\textbf{k}_{0}}(t)\rangle,
\end{align}
where $\alpha_{m,\textbf{k}_{0}}(t)$ is the complex amplitude for the electron in different energy bands. Following the idea of quantum path integral \cite{Feynman}, the evolution of the wave packet can be described in terms of a series of quantum paths with different sub-wave packets. The term $|\varphi_{m,\textbf{k}_{0}}(t)\rangle$ describes the dynamics of the sub-wave packet in the \textit{m}-th band. The corresponding wave-vector $\textbf{k}(t)$ of the Bloch state is denoted as the trajectory in reciprocal space. The term $e^{-i\int_{t_{0}}^{t}d\tau E_{m}(\textbf{k}(\tau))}$ describes the dynamical phase accumulated during the evolution, and the term $\alpha_{m,\textbf{k}_{0}}$ describes the weight factor. In the framework of RST, one can obtain a trajectory perspective on HHG in solid as illustrated in Fig. \ref{fig2}. Without loss of generality, here we consider only two bands (i.e., $m=c,v$ for the CB and VB), which is shown to work well in previous works \cite{Vampa2014,Vampa2015,Osika2017}. For an electron initially located at $\textbf{k}_{0}$ in VB:

(1) it can be tunnel ionized from the VB to the CB at time $t'$ with certain probability (denoted by the purple dashed arrows at ``$t'_{1}$'' and ``$t'_{2}$''). Before ionization, the electron oscillates in the VB. After ionization, the trajectory will split into two parts (denoted by the white dashed lines on the VB and CB): one corresponds to the electron remaining and oscillating in the VB; the other one corresponds to the electron oscillating in the CB. When the depletion of the initial state can be neglected, one can obtain the complex amplitude of the ionization rate $\chi_{\textbf{k}_{0}}(t') = e^{iS_{c,\textbf{k}_{0}}(t',t_0)}\Omega_{cv}(t')e^{-iS_{v,\textbf{k}_{0}}(t',t_0)}$ \cite{SM}. $\Omega_{cv}(t) = \vec{\xi}_{cv}(\textbf{k}(t))\cdot \textbf{F}(t)$ describes the amplitude of the transition probability, where the transition matrix elements $\vec{\xi}_{cv}(\textbf{k}) = \langle c,\textbf{k}|\nabla_{\textbf{k}}|v,\textbf{k}\rangle_{cell} = -i\langle c,\textbf{k}|\hat{\textbf{r}}|v,\textbf{k}\rangle = -i\textbf{d}_{cv}(\textbf{k})$. The relative phase includes two parts: one is the phase $S_{v,\textbf{k}_{0}}(t',t_{0}) = \int_{t_{0}}^{t'}d\tau E_{v}(\textbf{k}(\tau))$ accumulated in the VB; the other one is the phase $S_{c,\textbf{k}_{0}}(t',t_{0}) = \int_{t_{0}}^{t'}d\tau E_{c}(\textbf{k}(\tau))$ accumulated in the CB.
Note that the trajectories of different ionization times are coherent with each other.

(2) During the oscillations, the electron is accelerated by the external field and the trajectory in reciprocal space $\textbf{k}(t)$ satisfies the acceleration theorem. The electron gains energy $E_{c}(\textbf{k}(t))$ and $E_{v}(\textbf{k}(t))$ in the CB and VB, respectively. Their energy difference is denoted by $\Delta E(\textbf{k}(t))=E_{c}(\textbf{k}(t))-E_{v}(\textbf{k}(t))$.


(3) During the above process in the external field, two kinds of induced currents are produced \cite{Vampa2014}. An intraband current is induced by the charge transfer inside the VB and the CB. An interband current is induced by the polarization between the electron in the VB and CB. The induced intraband and interband currents are expressed as:
\begin{align}\label{currents}
  &j_{mm,\textbf{k}_{0}}(t) = \textbf{p}_{mm}(\textbf{k}(t)), \nonumber \\
  j_{cv,\textbf{k}_{0}}(t) &= e^{iS_{c,\textbf{k}_{0}}(t,t_{0})}\textbf{p}_{cv}(\textbf{k}(t))e^{-iS_{v,\textbf{k}_{0}}(t,t_{0})},
\end{align}
where $\textbf{p}_{mm}(\textbf{k}) = \langle\phi_{m,\textbf{k}}|\hat{\textbf{p}}|\phi_{m,\textbf{k}}\rangle = \nabla_{\textbf{k}}E_{m}(\textbf{k})$, $\textbf{p}_{cv}(\textbf{k}) = \langle\phi_{c,\textbf{k}}|\hat{\textbf{p}}|\phi_{v,\textbf{k}}\rangle = -i\textbf{d}_{cv}(E_{c}(\textbf{k})-E_{v}(\textbf{k})) = -i\textbf{d}_{cv}\Delta E(\textbf{k})$. 

As in general semiconductors, the VB is fully occupied. The trajectories from different initial states and different ionization times form a trajectory-ensemble. In this case, the total currents can be expressed as $J(t) = J^{intra}(t)+J^{inter}(t)$ with \cite{SM}
\begin{align}\label{Scurrents}
  J^{intra}(t) &= \sum_{m=c,v}\sum_{\textbf{k}_{0}\in \textrm{BZ}}|\sum_{t'}\alpha_{m,\textbf{k}_{0}}(t,t')|^2\textbf{j}_{mm,\textbf{k}_{0}}(t),  \nonumber \\
  J^{inter}(t) &= \sum_{\textbf{k}_{0}\in \textrm{BZ}}\sum_{t'}\alpha^{*}_{c,\textbf{k}_{0}}(t,t')\textbf{j}_{cv,\textbf{k}_{0}}(t)+c.c.,
\end{align}
where $\alpha_{c,k_0}(t,t')=\chi_{\textbf{k}_{0}}(t')\theta(t,t')$ and $\alpha_{v,k_0}(t,t')$, satisfying $|\alpha_{v,k_0}(t,t')|^2=1-|\alpha_{c,k_0}(t,t')|^2$, describe the weight of the electron trajectories initially located at $\textbf{k}_0$ and ionized at time $t'$. $\theta(t,t')$ is a step function with $\theta(t,t')=0$ for $t< t'$ and $\theta(t,t')=1$ for $t\geq t'$, which denotes the occurrence of ionization at $t'$. The HHG spectrum can be obtained by the Fourier transform of the currents.


\begin{figure}[!t]
  \includegraphics[width=10cm]{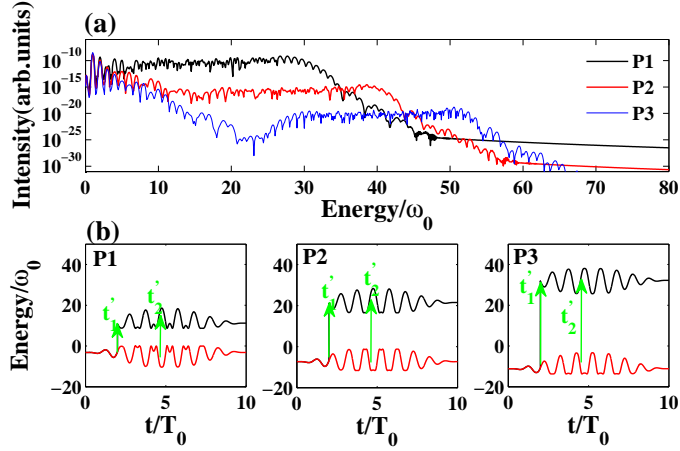}
  \caption{(a) Harmonic spectra with linear field and (b) the corresponding trajectories from the electron initially located at $P1(0.15,0)$, $P2(0.24,0.24)$ and $P3(0.30,0.43)$. For clarify, we chose the trajectories from the ionization time $t'_{1}$ and $t'_{2}$. The laser parameters are the same as that in Fig. 2.}\label{fig3}
\end{figure}

To validate the above model, we calculate the HHG spectra from ZnO under the same conditions as in Fig. \ref{fig1}. It is shown that the HHG spectra obtained with RST method are almost identical with those from the SBE simulations (see the Supplementary Material \cite{SM}). More importantly, RST model enables one to anatomize the HHG contributed by different trajectories as well as how these trajectories change the structure of the harmonic spectra. According to Eq. \ref{Scurrents}, one can obtain the high harmonics $Y_{\textbf{k}_{0}}(\omega)$ for different trajectories by the Fourier transform of the current $(\sum_{m=c,v}|\sum_{t'}\alpha_{m,\textbf{k}_{0}}(t,t')|^2\textbf{j}_{mm,\textbf{k}_{0}}(t))+(\sum_{t'}\alpha^{*}_{c,\textbf{k}_{0}}(t,t')\textbf{j}_{cv,\textbf{k}_{0}}(t)+c.c.)$. We first discuss the HHG in the linearly polarized field. By analyzing the HHG contributed by different trajectories, one can find the link between the high harmonics and the trajectories from different initial states. Figure \ref{fig3} shows three representative trajectories and the generated harmonic spectra by these trajectories. The harmonic spectra are denoted as P1, P2 and P3, and the corresponding electrons are initially located at $\textbf{k}_{0}=(0.15,0)$ a.u., $\textbf{k}_{0}=(0.24,0.24)$ a.u. and $\textbf{k}_{0}=(0.30,0.43)$ a.u., respectively. The red and black lines in Fig. \ref{fig3}(b) shows the trajectories of electron oscillating in the VB and CB, respectively. The ionization can occur at different times, for example at $t'_1$ and $t'_2$ as denoted by the green vertical arrows. For P1, during the oscillation, the energy difference $\Delta E(\mathbf{k_0}+\mathbf{A}(t))$ gained by the electrons in the VB and CB is in the range from $10\omega_{0}$ to $29\omega_{0}$, agreeing well with the spectral range of the obtained high harmonics as shown in Fig. \ref{fig3}(a). One can also find the good correspondence between the high harmonics and trajectories P2 and P3. According to the generated photon energy, we identify three kinds of trajectories in the reciprocal-space: the trajectories contributing (I) only to the first plateau ranging from $10\omega_{0}$ to $29\omega_{0}$, (II) only to the second plateau ranging from  $40\omega_{0}$ to $55\omega_{0}$, and (III) to both plateaus. The three representative trajectories P1, P2, and P3 shown in Fig. \ref{fig3} right corresponds to the cases (I), (II), and (III), respectively.

\begin{figure}[!t]
	  \includegraphics[width=10cm]{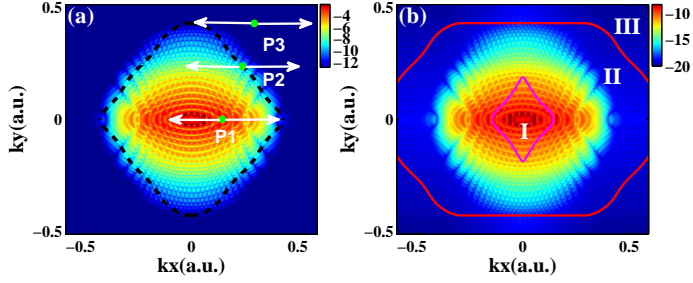}
  \caption{ (a) Real population and (b) harmonic yield contributed by the trajectories of electrons initially located at different points in the reciprocal-space. The points mark the positions of P1, P2 and P3 in Fig. 3, and the white arrows mark the oscillation range of the corresponding trajectories.}\label{fig4}
\end{figure}

Further, we analyze the harmonic yield contributed by the trajectories of electrons located at different initial momenta $\textbf{k}_0$. To this end, we average the intensity of the HHG contributed by trajectories of electron initially located at $\textbf{k}_{0}$, i.e.,
\begin{align}\label{Averagingyield}
  \overline{Y}_{\textbf{k}_{0}} = \frac{\int_{\min[\Delta E(\textbf{k}_{0}+\textbf{A}(t))]}^{\max[\Delta E(\textbf{k}_{0}+\textbf{A}(t))]}|Y_{\textbf{k}_{0}}(\omega)|^{2}d\omega}{\max[\Delta E(\textbf{k}_{0}+\textbf{A}(t))]-\min[\Delta E(\textbf{k}_{0}+\textbf{A}(t))]}.
\end{align}
The result is shown in Fig. \ref{fig4}(b). In Fig. \ref{fig4}(a), we plot electron population by coherently sum up the weight of electron trajectories initially located at $\textbf{k}_{0}$, which is calculated as $|\int_{t_0}^{\infty}\alpha_{c,\textbf{k}_{0}}(t,t')dt'|^2$. This is identified as the real population in \cite{Krausz2014}. Since we discuss the trajectories with different crystal momenta separately, we call it $\textbf{k}$-dependent real population.
By comparing Figs. \ref{fig4}(a) and (b), one can see that both the $\textbf{k}$-dependent real population and the harmonic yield exhibit the similar interference structure, which is due to the interference of different trajectories. A similar interference also can be seen in photoelectron holography of gaseous medium \cite{Bian2011}. Besides the similar structure, the relative intensity of the harmonics is also proportional to that of the real population. Therefore, one can understand the HHG process by analyzing $\textbf{k}$-dependent real populations.

The black dashed curve in Fig. \ref{fig4}(a) marks the positions satisfying $\Delta E(\textbf{k}) = 29\omega_{0}$, which corresponds to the cutoff energy of the first plateau. 
Based on the discussions about Fig. \ref{fig2}, three parts in the BZ can be identified. Part I is inside the purple curve (near the centre of the BZ i.e., the top of the VB), from which the trajectories satisfy $\textrm{max}[\Delta E(\textbf{k}_{0}+\textbf{A}(t))]\leq 29\omega_{0}$. The trajectories with initial $\textbf{k}_{0}$ in this part are totally inside the black dashed curve and contribute only to the first plateau (see P1 in Fig. \ref{fig4} (a)). Part III is outside the red curve (far away from the centre of the BZ), from which the trajectories satisfy $\textrm{min}[\Delta E(\textbf{k}_{0}+\textbf{A}(t))]\geq 29\omega_{0}$ and contribute only to the second plateau (see P3 in Fig. \ref{fig4} (a)). Part II is the region between Parts I and III. As shown in Fig. \ref{fig4} (a), the real population in Part I is much higher than that of Part III and this leads to the two-plateau structure in the linearly polarized field.



\begin{figure}[!t]
  \includegraphics[width=10cm]{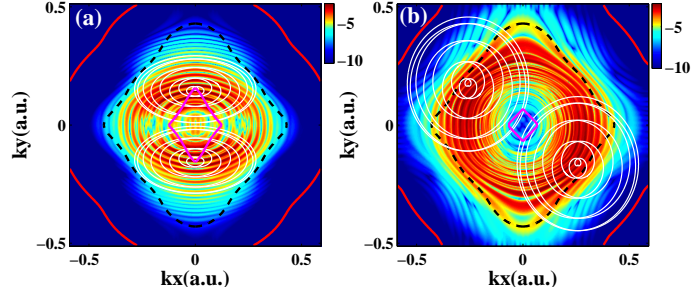}
  \caption{ The \textbf{k}-dependent real populations for ellipticity (a) $\varepsilon=0.5$ and (b) $\varepsilon=1$. The black dashed curves mark the positions where $E_{c}(\textbf{k})-E_{v}(\textbf{k}) = 29\omega_{0}$. The white solid lines mark the trajectories from two initial conditions.}\label{fig5}
\end{figure}


Next, we come to the abnormal ellipticity dependencies of the HHG in the first and second plateaus. We consider two ellipticity values $\varepsilon=0.5$ and $\varepsilon=1$. As in linear polarization, three parts in the BZ can also be identified for elliptical and circular polarizations. The borders of these parts are marked by purple and red curves in Fig. \ref{fig5}. First, we discuss the case of $\varepsilon=0.5$. As shown in Fig. \ref{fig5}(a), the $\textbf{k}$-dependent real population also shows an interference structure, but the interference fringe shows an elliptical ringlike structure and two dominant areas appear on the $k_y$ axis. Two representative trajectories from the dominant areas are plotted by the white lines in Fig. \ref{fig5}(a). The oscillation ranges of these trajectories are all within the black dashed curve. In other words, these trajectories only contribute to the first plateau. The real population decreases dramatically out of the dominant areas and hence the harmonic yield in the first plateau is still much higher than that of the second plateau. Then, we discuss the case of $\varepsilon=1$. As is shown in Fig. \ref{fig5}(b), the \textbf{k}-dependent real population shows a donut structure with the dominant distribution at about $|\textbf{k}| = A_0= F_{0}/\omega$, where $A_0$ and $F_{0}$ are the amplitudes of the vector potential and electric field. The oscillating trajectories (marked by the white lines) cover both regions inside and outside the black dashed curve. Then the two-plateau structure is smeared out. Moreover, the \textbf{k}-dependent real population in the elliptically and circularly polarized field shows a hollow structure near the top of VB ($\textbf{k}=(0,0)$ a.u.) and the hollow area is larger with increasing the ellipticity. This is similar to the photoionization of gases in elliptically and circularly polarized fields, where the electron is ionized with non-zero vector potential. Due to this hollow structure, one can see that the trajectories with large real populations dominantly contribute to the high energy regions of HHG. This explains why the first plateau is suppressed and the second plateau becomes stronger with increasing the ellipticity.

As shown in Fig. \ref{fig4} and Fig. \ref{fig5}, the real population shows a minim at the top of the VB ($|\textbf{k}|=0$ a.u.) and a maximum around $|\textbf{k}|=0.07$ a.u. (linear polarization), $|\textbf{k}|=0.15$ a.u. (elliptical polarization) and $|\textbf{k}|=0.28$ a.u. (circular polarization). This clearly indicates that the electron trajectories initially deviating from the top of the VB play an important role. By considering these trajectories, the cutoff energy of the first plateau becomes $\textrm{max}[\Delta E(\textbf{k}_{max}+\textbf{A}(t))]$ ($\textbf{k}_{max}$ denotes the crystal momenta with the maximum real population in Figs. \ref{fig4} and \ref{fig5}). In contrast, it is previously assumed that the electron initially at the top of the VB is dominant \cite{MW2015,Bian2017}. In this case, the generated harmonic cutoff is predicted to be $\textrm{max}[\Delta E(\textbf{A}(t))]$, which is smaller than that from the SBE simulation. Instead, the new cutoff $\textrm{max}[\Delta E(\textbf{k}_{max}+\textbf{A}(t))]$ agrees well with the SBE simulation. This also explains the extension of cutoff energy in the circularly polarized field.

In conclusion, we investigate the HHG in solids driven by laser fields with different ellipticities. We find a two-plateau structure in the spectrum and distinct ellipticity dependencies of the HHG in the first and second plateaus. To understand these phenomena, we develop the RST method, which builds a bridge between the HHG yield and the microscopic dynamical processes of electrons. With this method, the two-plateau structure and the abnormal ellipticity dependencies are quantitatively and intuitively explained in terms of electron trajectories. It is shown that the trajectories initially deviating from the top of the VB play a very important role, which leads to a larger harmonic cutoff energy. Moreover, because the real populations exhibit a donut distribution in the circularly polarized field, the trajectories initially deviating from the top of the VB dramatically enhance the HHG in the second plateau. Our method provides not only a quantitative description but also a clear picture the HHG in solids. It will be a useful tool to understand the HHG process, and it can be extended for more complicated forms of driving fields by analyzing the trajectory ensembles with similar procedures. This will shed light on the manipulation and optimization the HHG in solids with a well-designed driving field. 

\begin{acknowledgments}
We gratefully acknowledge G. Vampa for valuable discussions.
This work was supported by the National Natural Science Foundation of China
under Grants No. 11627809, 61475055, 11704137 and 11404123. Numerical
simulations presented in this paper were carried out using the High Performance
Computing experimental testbed in SCTS/CGCL (see http://grid.hust.edu.cn/hpcc).
\end{acknowledgments}

\end{document}